\documentclass[fleqn,12pt,twoside]{article}
\usepackage[headings]{espcrc1}
\usepackage{graphicx,wrapft,amsmath,amssymb,epsfig}

\title{Hadron Physics at J-PARC}

\author{S. Kumano
\address{Institute of Particle and Nuclear Studies, KEK 
         and Department of Particle and \\ Nuclear Studies,
         Graduate University for Advanced Studies \\
         1-1, Ooho, Tsukuba, 305-0801, Japan}}

\begin{document}

\maketitle

\begin{abstract}
An outline is explained for hadron-physics projects at J-PARC, which
is considered to be one of the flagship facilities in hadron 
physics from 2008. The facility provides an intensity frontier with
50 GeV proton beam for nuclear and particle physics. 
It could cover a wide range of hadron physics from strongly
interacting many-body systems with an extended hadronic degree
of freedom, strangeness, to new forms of hadrons and hadronic matters.
These studies lead not only to create innovative fields of
hadron physics but also, possibly, to understand fundamental
interactions because of recent progress on AdS/CFT correspondence. 
At the first stage of the J-PARC operation, hadron topics are mainly
on strangeness nuclear physics such as hypernuclei, kaonic nuclei, and
possible pentaquark hadrons. Then, the studies could be extended to
exotic hadron searches, chiral dynamics in nuclear medium, structure
functions, hard exclusive processes, hadron physics in neutrino
scattering, and spin structure of the nucleon.
With major upgrades of the facility, extensive studies could be done
for the nucleon spin, heavy-ion physics, and hadron physics
at a high-energy neutrino factory. 
\end{abstract}

%%%%%%%%%%%%%%%%%%%%%%%%%%%%%%%%%%%%%%%%%%%%%%%%%%%%%%%%%%%%%%%%%%%%%%%%%%%%%%%%
\section{Introduction}

Nuclear physics has a long history of investigations starting from
nucleon-nucleon ($NN$) potentials which are determined from many $NN$
scattering measurements and deuteron properties. From these studies,
ordinary nuclei have been understood in details. Now, it is time to
step into a realm of new hadronic matters, which do not usually exist
in nature, by extending isospin and flavor degrees of freedom.
These are the fields of unstable nuclei and strangeness nuclear physics,
and they have important applications to the studies of evolution of
our universe, particularly in nucleosynthesis, supernova explosion, 
and neutron stars. Another direction of nuclear physics is to create
extraordinary materials by changing density and temperature, and
then to investigate their properties.

Hadron structure has been also investigated extensively, and gross
properties are now understood. However, it is unfortunate that
one of basic physical quantities, the proton spin, has not been
clarified yet in a fundamental level from quarks and gluons.
Now, it is time to investigate further the basic properties like
the nucleon spin, then to explore innovative fields on new forms of
hadronic matters. In this way, modern hadron and nuclear physics
is considered as a challenging field on
(1) description of hadrons and nuclei by the quark and gluon degrees
    of freedom and
(2) new quantum many-body systems in extreme conditions
    by changing isospin, flavor, density, and temperature. 

Many of these topics could be covered at the Japan Proton Accelerator Research
Complex (J-PARC) \cite{jparc,jparc-loi,jparc-theory} which will be completed
in 2008. At the first stage, strangeness nuclear physics, particularly
hypernuclear physics, kaonic nuclei, and pentaquarks, are
investigated together with neutrino-oscillation physics. The neutrino
experiment is also related to hadron physics because accurate description
of neutrino interactions with the target water requires detailed
knowledge of hadron and nuclear physics.
At the second stage, other hadron-physics topics could be investigated
\cite{j-parc-hs05}. They include studies on exotic hadrons, chiral symmetry
and meson properties in a nuclear medium, structure functions, 
hard exclusive processes, and spin physics with target polarization.
After major upgrades: proton-beam polarization, heavy-ion beam,
and high-energy neutrino factory, much detailed investigations will become
possible for spin physics, high-density matters, and hadron structure
with an intense high-energy neutrino beam. An outline of these topics
is explained in this paper.

We would like to stress that the J-PARC hadron physics is intended
to explore diversity of hadrons and their bound systems by investigating
properties of hadronic interactions and new matters. These studies are not
explicitly aimed at finding new elementary interactions as particle
physicists do. However, such studies may shed light on fundamental
interactions themselves because some observables in hadron physics are
possibly related to quantities in string theory because of a recent
progress on duality \cite{sjb-06}.

%%%%%%%%%%%%%%%%%%%%%%%%%%%%%%%%%%%%%%%%%%%%%%%%%%%%%%%%%%%%%%%%%%%%%%%%%%%%%%%%
\section{Strangeness Nuclear Physics}

Strangeness nuclear physics is investigated at the first stage of
J-PARC by using secondary kaon and pion beams. There are four
important points in the strangeness physics. 

First, strange baryons are not affected by the Pauli exclusion principle,
so that they could penetrate into deep inside of a nucleus. 
They should be good probes for nuclear medium.

Second, the strange-quark mass is a special one.
The up- and down-quark masses are much smaller than the QCD
(Quantum Chromodynamics) scale parameter $\Lambda$, whereas the charm-
and bottom-quark masses are much larger. It means that chiral symmetry
is a guiding principle in the up- and down-quark mass region, and
nonrelativistic quark models should work in the high-mass region.
On the other hand, the strange-quark mass is of the same order of
$\Lambda$, which means that it is rather difficult to describe
hadrons with strangeness. However, it could be viewed as an advantage
in the sense that the strange quark could be an appropriate quantity
to probe QCD dynamics.

\begin{wrapfigure}{r}{0.38\textwidth}
\vspace{-1.2cm}
\begin{center}
     \includegraphics[width=0.37\textwidth]{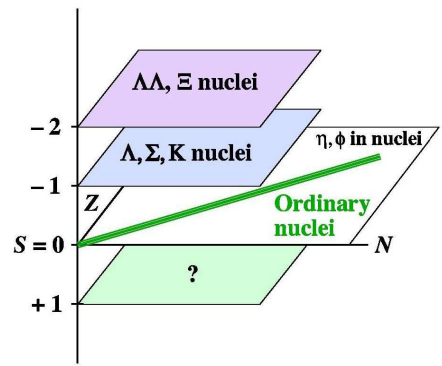}
\end{center}
\vspace{-0.8cm}
\caption{Nuclei with strangeness.}
\vspace{-0.6cm}
\label{fig:strangenuclei}
\end{wrapfigure}

Third, new hadronic many-body systems with strangeness can be created.
In contrast to ordinary nuclei, which are described with well known
$NN$ potentials from many scattering data, only a few dozen data
exist for determining hyperon-nucleon ($YN$) interactions. We inevitably
have large theoretical uncertainties in describing hypernuclei. Instead
of building $YN$ potentials only from the scattering data, the potentials
should be deduced also from properties of hyper- and kaonic nuclei.
As shown in Fig.\ref{fig:strangenuclei}, ordinary nuclei exist
in the $S=0$ plane, where $S$ is the strangeness.
At J-PARC, new forms of nuclei are investigated by extending
the flavor degree of freedom. The strangeness $-1$ and $-2$ nuclei are
measured by the reactions, $(K^-,\pi^\pm)$, $(\pi^\pm,K^+)$, and
$(K^-,K^+)$. In addition, \\
%\vspace{-0.5cm}
%%%%%%%%%%%%%%%%%%%%%%%%%%%%%%% figure %%%%%%%%%%%%%%%%%%%%%%%%%%%%%%%
%%%%%%%%%%%%%%%%%%%%%%%%%%%%%%% figure %%%%%%%%%%%%%%%%%%%%%%%%%%%%%%%
\noindent
strangeness $+1$ nuclei may be investigated if a pentaquark hadron
exists. 

Fourth, the strange quark could be considered an impurity in hadronic
materials. There is an interesting theoretical proposal for  
a new form of strangeness hadrons.  For example, $K^-$ forms
a deeply bound state with $^3$He, and it is called a strange tribaryon.
Theoretically, the $K^- N$ potential was determined from
$\overline K N$ scattering data, the $1s$ state of the kaonic hydrogen atom,
and $\Lambda (1405)$ as a $\overline K N$ bound state, and then such strange
tribaryons are studied. The kaon plays a role of shrinking the nucleus, and 
the density becomes significantly higher \cite{trib}. If this theoretical
estimation is correct, a new region of hadronic phase diagram, a high-density
and low-temperature region, could be explored by studying these hadrons.
The strange tribaryons were experimentally investigated in the reaction
$^4$He(stopped $K^-,N)$, where $N$=proton or neutron, at KEK-PS by
using a $K^-$ beam \cite{kek-ps-e471}. In missing mass spectra,
they reported bumps which could be bound states of $K^-pnn$ and $K^-ppn$.
There is also an interesting report on a possible $K^-pp$ bound
state by the FINUDA collaboration \cite{finuda}.
We should note that there is a critical view on the theoretical calculations
and the experimental data \cite{ot06}. Because these experiments are not
confirmed independently, we need further experimental efforts for their
verification. However, these studies could lead to a new field of
exotic-hadron physics by detailed studies at J-PARC.

%%%%%%%%%%%%%%%%%%%%%%%%%%%%%%%%%%%%%%%%%%%%%%%%%%%%%%%%%%%%%%%%%%%%%%%%%%%%%%%%
\section{Hadron spectroscopy and chiral dynamics}

%%%%%%%%%%%%%%%%%%%%%%%%%%%%%%%%%%%%%%%%%%%%%%%%%%%%%%%%%%%%%%%%%%%%%%%%%%%%%%%%
\subsection{Exotic Hadrons}

Basic properties of hadrons are described by the simple quark model with
$q\bar q$ and $qqq$ configurations. However, there are theoretical predictions
on exotic hadrons which cannot be described by these configurations.
They include tetraquarks ($q^2\bar q^2$), pentaquarks ($q^4\bar q$),
dibaryons ($q^6$), strange tribaryons ($q^{10}\bar q$), and glueballs ($gg$).
However, there is no undoubtable experimental evidence for the exotic structure. 
Recently, there are significant experimental contributions to exotic-hadron
search from Japanese facilities, although some of them are very controversial.
First, a pentaquark $\Theta^+$(1540) was reported at SPring-8. There are
measurements on the strange tribaryons $S^0(3115)$ and $S^+(3140)$ at KEK-PS,
possible candidates for tetraquarks or $D\bar D$ molecules $X(3872)$
and $Y(3940)$ at KEK-Belle, and tetraquarks or $DK$-molecule candidates
$D_{sJ}(2317)$ and $D_{sJ}(2460)$. Because of the upsurge of these
experimental findings, we think that time has come to redescribe
the basic hadron structure with exotic quark and gluon configurations.

The existence of the pentaquark $\Theta^+$ has been heatedly debated 
for a few years. Although there are a number of negative experiments,
it is still not very clear whether or not it actually exists. Obviously,
a clear decisive experiment is needed to conclude its identification. 
There is a proposal to investigate $\Theta^+$ at the first stage of J-PARC
by using the reaction $\pi^- +p \rightarrow K^- +X$ \cite{j-parc-theta}.
The $\pi^-$ beams with 1.87, 1.92, and 1.97 GeV/c will be used
in the K1.8 beamline. A more decisive test on the $\Theta^+$ existence is
considered at the next stage \cite{j-parc-theta}. Using a low-momentum
(475 MeV/c) $K^+$ beam at J-PARC, they will investigate an $s$-channel
formation, $K^+ +n \rightarrow \Theta^+ \rightarrow K^0 +p$. This 
formation should be a crucial experiment on the existence of $\Theta^+$.
Because there are many theoretical predictions on exotic hadrons 
in addition to the pentaquark, experiments will be continued to search
for other exotic ones.

%%%%%%%%%%%%%%%%%%%%%%%%%%%%%%%%%%%%%%%%%%%%%%%%%%%%%%%%%%%%%%%%%%%%%%%%%%%%%%%%
\subsection{Chiral Dynamics in Nuclear Medium}

%%%%%%%%%%%%%%%%%%%%%%%%%%%%%%% figure %%%%%%%%%%%%%%%%%%%%%%%%%%%%%%%
\begin{wrapfigure}{r}{0.35\textwidth}
\vspace{-1.5cm}
\begin{center}
     \includegraphics[width=0.33\textwidth]{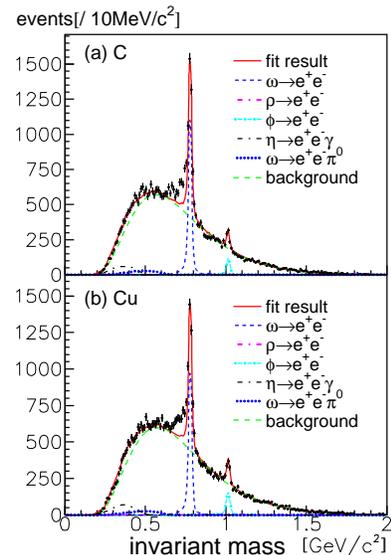}
\end{center}
\vspace{-0.7cm}
\caption{KEK-E325 experiment (from \cite{kek-ps-e325}).}
\vspace{-0.7cm}
\label{fig:e325}
\end{wrapfigure}
%%%%%%%%%%%%%%%%%%%%%%%%%%%%%%% figure %%%%%%%%%%%%%%%%%%%%%%%%%%%%%%%

Current masses of up- and down-quarks are very small in comparison with
the nucleon mass, which makes us wonder what the mechanism is to 
generate light hadron masses. It could be explained by
chiral-symmetry breaking, which leads to a nonvanishing expectation
value $< q\bar q> \ne 0$. This quantity is called scalar quark condensate
and it is an order parameter for the chiral phase transition.
Because it is not a direct physical observable, we need to find
experiments for testing the idea.

One of such experiments is to measure vector-meson masses in a nucleus.
There are theoretical predictions on modifications
of their masses due to partial restoration of chiral symmetry
inside a nuclear medium \cite{mass-mod}. Typical effects are about 18\%
reduction in $\rho$ and $\omega$ masses. There are measurements
on these masses. For example, recent KEK-E325 experimental results are
shown in Fig.\ref{fig:e325}. Using the 12 GeV proton beam at the KEK-PS
facility, they measured the reactions $p+A \rightarrow \rho,\ \omega,\ \phi+X$ 
($\rho,\ \omega,\ \phi \rightarrow e^+ +e^-$) \cite{kek-ps-e325}.
Fitting the data by a density-dependent mass, $m(\rho)/m(0)=1-k \rho/\rho_0$,
for $\rho$ and $\omega$, they obtained a 9\% mass shift
($k=0.092 \pm 0.002$). It indicates a clear signature for the existence
of the mass shift. At J-PARC, this kind of studies will be continued
for accurate determination of the mass shifts.
 
%%%%%%%%%%%%%%%%%%%%%%%%%%%%%%%%%%%%%%%%%%%%%%%%%%%%%%%%%%%%%%%%%%%%%%%%%%%%%%%%
\section{Structure Functions and Hard Exclusive Processes}

%%%%%%%%%%%%%%%%%%%%%%%%%%%%%%%%%%%%%%%%%%%%%%%%%%%%%%%%%%%%%%%%%%%%%%%%%%%%%%%%
\subsection{Structure Functions}\label{sec:sf}

Parton distribution functions (PDFs) could be investigated at J-PARC.
In the Drell-Yan process, kinematical variables are related by
$x_1 x_2=m_{\mu\mu}^2/s$ with the parton momentum fractions $x_1$
and $x_2$, center-of-mass energy $\sqrt{s}$, and dimuon mass $m_{\mu\mu}$.
Roughly speaking, the probed $x$ region is given by
$x \sim m_{\mu\mu} /\sqrt{s}$. At LHC, the small-$x$ region 
($x \sim 10^{-5}$) is investigated, whereas the J-PARC is
an appropriate facility to investigate the large-$x$ region ($x>0.2$).
Because theoretical descriptions of hard processes could depend
much on effects from higher orders of perturbative QCD (pQCD)
and resummations  \cite{pqcd-jparc}
at the J-PARC energy $E_p$=50 GeV (30 GeV in the beginning),
such effects should be estimated for extracting information from
cross-section measurements. On the other hand, the J-PARC facility
could be viewed as an appropriate one for investigating the details
of pQCD and resummation physics. 

%%%%%%%%%%%%%%%%%%%%%%%%%%%%%%% figure %%%%%%%%%%%%%%%%%%%%%%%%%%%%%%%
\begin{wrapfigure}{r}{0.33\textwidth}
\vspace{-0.8cm}
\begin{center}
     \includegraphics[width=0.306\textwidth]{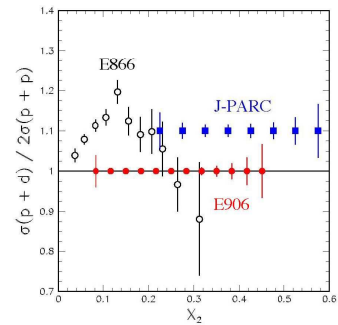}
\end{center}
\vspace{-0.9cm}
\caption{Drell-Yan experiment for $\bar d/\bar u$ \cite{jparc-dy}.}
\label{fig:jpark-dy}
\vspace{-0.6cm}
\end{wrapfigure}
%%%%%%%%%%%%%%%%%%%%%%%%%%%%%%% figure %%%%%%%%%%%%%%%%%%%%%%%%%%%%%%%

As one of the J-PARC experiments on structure functions, Drell-Yan
experiments are proposed \cite{jparc-dy}. For example, measuring
the cross-section ratio $\sigma_{DY}(p+d)$ $/2\sigma_{DY}(p+p)$, they
expect to extract the ratio $\bar d/\bar u$ in the nucleon
at $x>0.2$. This region has not been measured by the Fermilab-E866
experiment as shown in Fig.\ref{fig:jpark-dy}.
There is a perturbative QCD contribution on $Q^2$ evolution
of $\bar u-\bar d$ due to a $q \rightarrow \bar q$ splitting process
in the next-to-leading order \cite{sk-pr}. However, it is expected
to be small as long as the evolution is done in a perturbative QCD
region. The dominant source of a finite $\bar u-\bar d$
distribution comes from a nonperturbative mechanism such as meson
clouds in the nucleon \cite{sk-pr}. The measurements could shed light
on nontrivial ``peripheral structure" of the nucleon.

Nuclear PDFs could be also investigated in the large-$x$ region.
Modifications of valence-quark, antiquark, and gluon distributions and
their uncertainties are determined in Ref. \cite{npdf04}
by a global analysis of high-energy nuclear data.
Although the valence-quark distributions are well determined, the antiquark
and gluon distributions have large uncertainty bands at large $x$.
For applications to heavy-ion and neutrino-nucleus reactions, they should
be determined in a wide-$x$ region. The J-PARC facility could contribute
to the determination of the antiquark distributions at $x>0.2$ by
the Drell-Yan experiment.

%%%%%%%%%%%%%%%%%%%%%%%%%%%%%%%%%%%%%%%%%%%%%%%%%%%%%%%%%%%%%%%%%%%%%%%%%%%%%%%%
\subsection{High-energy spin physics with target polarization}

Search for the origin of the nucleon spin is one of important topics
in hadron physics. The proton-beam polarization requires a major upgrade
at J-PARC, so that spin physics should be investigated first only with
target polarizations.

There are recent studies on transverse single spin asymmetries (SSAs),
to which three mechanisms contribute: Sivers effect, Collins effect, and
higher twists \cite{ssa}. The Sivers effect is on unpolarized quark
distributions in the transversely polarized nucleon, whereas the Collins
effect is on fragmentation of polarized quark into unpolarized hadron.
The Sivers effect is especially interesting because it 
probes orbital angular momenta of quarks and gluons. Note that the nucleon
spin is carried by the orbital angular momenta in addition to
the quark and gluon spins. As an interesting SSA experiment
at J-PARC, $D$-meson production could be investigated \cite{goto}.
The process is important for extracting the Sivers effect
since the Collins mechanism does not contribute due to no charm
polarization. At the J-PARC energy, the asymmetry is sensitive to
the quark Sivers effect at $x_F<0$, whereas it is sensitive to
the gluon Sivers at RHIC.

The tensor structure function $b_1$ could be investigated in polarized
proton-deuteron Drell-Yan processes \cite{b1dy}. This new function does
not exist in the spin-1/2 nucleon. It was recently measured by
the HERMES collaboration \cite{hermes-b1}. Antiquark tensor
polarizations could be determined at J-PARC. The tensor structure
reflects dynamical aspects inside a hadron, so that
it is an interesting new field in hadron spin physics.

%%%%%%%%%%%%%%%%%%%%%%%%%%%%%%%%%%%%%%%%%%%%%%%%%%%%%%%%%%%%%%%%%%%%%%%%%%%%%%%%
\subsection{Hard exclusive processes}

There are other physics possibilities with the primary proton beam
although actual proposals have not been submitted yet. They include
$pp$ elastic scattering, generalized PDFs, and color transparency. 

First, $pp$ elastic scattering could be investigated for testing 
constituent counting rule and a prediction by the AdS/CFT
correspondence \cite{sjb-06}. It is also an interesting topic to
study the transition from hadron degrees of freedom to quark-gluon ones.
According to the counting theory, the cross section behaves
$d \sigma(AB \rightarrow CD)/dt \sim s^{2-n} f(\theta_{c.m.})$, where $n$
is the number of constituents: $n=n_A+n_B+n_C+n_D$. The exclusive process
$\gamma p \rightarrow \pi^+ n$ is recently investigated by the JLab-E94-104
experiment \cite{jlab-e94}, and experimental data clearly indicated
such a transition at $\sqrt{s} \sim$2 GeV and also the counting
rule at $\sqrt{s} >$2.5 GeV. A similar investigation could be done at J-PARC
for the $pp$ elastic scattering.

The second topic is on generalized parton distributions (GPDs).
Recently, there is a glowing interest in the studies of GPDs because
they should shed light on contributions from orbital angular momenta
to the nucleon spin. They were originally defined in virtual Compton
scattering; however, the definition of GPDs could be extended to 
transitions such as $N \rightarrow \Delta$ and $N \rightarrow \pi$
\cite{jparc-gpds}. Studies of the transition GPDs are initiated
recently, so that further theoretical investigations are needed about
their meanings and advantages.

The third is on color transparency by the process $pA \rightarrow pp (A-1)$
at J-PARC. At large momentum transfer, a small-size component
of the hadron wave functions should dominate a reaction cross section.
This small-size hadron could freely pass through nuclear
medium, so that it is called ``color transparency". It is 
a probe of dynamics in elementary reactions.
Nuclear transparency is defined by nucleonic and
nuclear cross sections: $T=\sigma_A/A\sigma_N$, which should increase
as the hadron size becomes smaller or the hard scale becomes larger.
The reaction $pA \rightarrow pp (A-1)$ was investigated by the BNL-EVA
experiment up to the energy 14 GeV \cite{bnl-eva}. A theoretical calculation
indicates a significant increase of the color transparency from 10 GeV
to 50 GeV in the J-PARC energy region \cite{jparc-color}, so that measurements
should provide important information on hadron interactions dynamics.

%%%%%%%%%%%%%%%%%%%%%%%%%%%%%%%%%%%%%%%%%%%%%%%%%%%%%%%%%%%%%%%%%%%%%%%%%%%%%%%%
\section{Hadron physics in neutrino scattering}

Neutrino-oscillation physics is one of the major projects from the first
stage of the J-PARC. Hadron and nuclear topics could be also investigated
in neutrino scattering.

First, nuclear corrections in the oxygen nucleus need to be taken into
account for accurate determination of the mixing angle $\theta_{13}$ because
a few percent accuracy is necessary for describing the cross section 
\cite{nuint}. The nuclear effects include binding, Fermi motion, Pauli
exclusion, $NN$ short-range correlations, and PDF modifications.
For example, a final-state nucleon suffers from the exclusion
due to the existence of other nucleons in a nucleus. The exclusion effects
produce a significant decrease of the cross section at small $Q^2$ where
the cross section is the largest. In the neutrino energy region,
$E_\nu \sim$1 GeV, this effect is estimated about 8\%. In addition,
the short-range $NN$ correlation gives rise to a large momentum tail
beyond the Fermi momentum, and it also produces modifications.
These effects should be carefully estimated for extracting accurate
information on the neutrino oscillation. There is another direction to
understand relatively low $Q^2$ reactions by using high-energy
information. Quark-hadron duality could be used 
for calculating neutrino cross sections as extrapolations from the 
deep inelastic region \cite{nuint}.

Second, strange-quark polarization can be determined by measuring
the axial-vector form factor in neutrino scattering \cite{delta-s}.
One of the issues for finding the origin of the nucleon spin is that
the antiquark polarization is not accurately determined.
This neutrino experiment could answer this issue. At J-PARC,
an off-axis (2.5$^\circ$) neutrino beam with relatively low energies
is ideal for the form-factor measurement with a near detector.
There are nonstrange and strange parts in the form factor. The nonstrange
one is fixed by the neutron $\beta$ decay at $Q^2$=0 and the strange one
becomes the strange-quark contribution to the nucleon spin:
$G_1^s (Q^2=0) = \Delta s$ at $Q^2 \rightarrow 0$. Liquid
scintillators with different mixture of
hydrogen and carbon are considered to remove nuclear effects in extracting
$\Delta s$. Expected experimental error of $\Delta s$ is about 0.03 which
is much smaller than the one by the BNL-E734 experiment.

%%%%%%%%%%%%%%%%%%%%%%%%%%%%%%%%%%%%%%%%%%%%%%%%%%%%%%%%%%%%%%%%%%%%%%%%%%%%%%%%
\section{Physics after major upgrades: Nucleon Spin, Heavy-Ion Physics,
         Hadron Physics at a Neutrino Factory}

As discussed in this paper, many important hadron topics will be
investigated with kaon, pion, neutrino, and proton beams. However,
much wider projects are covered by major upgrades of the facility.
They include the proton-beam polarization, heavy-ion beam, and 
neutrino factory. 

The polarized proton-proton reactions have been investigated recently
at RHIC (Relativistic Heavy Ion Collider)
and they started to obtain important information on the gluon
polarization. As explained in Sec. \ref{sec:sf}, the J-PARC is
a complementary facility to RHIC because a relatively large-$x$ region
can be investigated. Feasibility is studied for the proton polarization
at J-PARC, and we found that it is possible without technical difficulties
\cite{j-parc-pol}. The J-PARC could significantly contribute
to the clarification of the nucleon spin especially in the large-$x$ region.

The heavy-ion beam is another choice as a future J-PARC project. 
The RHIC and LHC (Large Hadron Collider) aim to investigate high-temperature
and high-density region of the hadron phase diagram. The J-PARC could
investigate lower-temperature and high density region, so that
it could be complementary to RHIC and LHC. In recent years, color
superconductivity has been studied extensively, so that
the low-temperature region becomes increasingly interesting. However,
the importance of the heavy-ion project at J-PARC should be examined
because similar heavy-ion topics could be investigated at GSI
in the same energy region.

There are feasibility studies for a neutrino factory with the energy 
of 30 GeV at J-PARC as well as the ones in Europe and US \cite{nu-fact}.
It could be used for various topics in hadron physics. In particular,
the energy is high enough to measure structure functions and the intensity
is high enough to have proton and deuteron targets, whereas a heavy iron
target is used in the CCFR and NuTeV experiments. It becomes possible
to obtain accurate valence-quark distributions in the ``nucleon"
by measuring $F_3$ without worrying about nuclear corrections.
In addition, if $F_3$ can be obtained for the deuteron, nuclear
modifications for the valence-quark distributions should be clarified
by $F_3^A/F_3^D$, especially at small $x$. There are new spin-dependent
structure functions, $g_3$, $g_4$, and $g_5$, in polarized neutrino-nucleon
scattering. From the function $g_5$, the polarization
of valence-quark distributions should become clear, whereas their magnitudes
are determined mainly by semi-leptonic decays at this stage. Furthermore,
the polarized singlet quark distribution is obtained by 
$g_1^{\nu p}+g_1^{\bar\nu p}$, which leads to clarification of quark-spin
content issue \cite{nu-fact}.

%%%%%%%%%%%%%%%%%%%%%%%%%%%%%%%%%%%%%%%%%%%%%%%%%%%%%%%%%%%%%%%%%%%%%%%%%%%%%%%%
\section{Summary}

The J-PARC facility can make significant contributions to a wide variety
of hadron topics starting from strange nuclear physics. 
At the second stage, other topics such as chiral dynamics, exotic hadrons,
structure functions, exclusive processes, and nucleon spin could be
investigated. There are also possibilities for major upgrades 
in order to investigate the details of nucleon spin, heavy-ion physics,
and hadron structure at a neutrino factory.

\vspace{-0.00cm}
%%%%%%%%%%%%%%%%%%%%%%%%%%%%%%%%%%%%%%%%%%%%%%%%%%%%%%%%%%%%%%%%%%%%%%%%%%%%%%%%

%%%%%%%%%%%%%%%%%%%%%%%%%%%%%%%%%%%%%%%%%%%%%%%%%%%%%%%%%%%%%%%%%%%%%%%%%%%%%%%%

\end{document}